\documentclass[aps,prl,floats,twocolumn,showpacs,superscriptaddress]{revtex4}

\usepackage{graphicx}
\usepackage{times}
\usepackage{graphics,dcolumn,bm,fleqn,epic,eepic,float}
\usepackage{amssymb,amsmath,multirow,rotate,color}
\usepackage{subfigure}
\begin{document}

\title{Effects of mobility in a population of Prisoner's Dilemma players}

\author{S. Meloni}
\affiliation{Department of Informatics and Automation, University of Rome "Roma Tre", Via della Vasca Navale, 79
00146, Rome, Italy}

\author{A. Buscarino}
\affiliation{Dipartimento di Ingegneria Elettrica, Elettronica e dei Sistemi,
Universit\`a degli Studi di Catania, viale A. Doria 6, 95125 Catania, Italy}
\affiliation{Laboratorio sui Sistemi Complessi, Scuola Superiore
di Catania, Via San Nullo 5/i, 95123 Catania, Italy}

\author{L. Fortuna}
\affiliation{Dipartimento di Ingegneria Elettrica, Elettronica e dei Sistemi,
Universit\`a degli Studi di Catania, viale A. Doria 6, 95125 Catania, Italy}
\affiliation{Laboratorio sui Sistemi Complessi, Scuola Superiore
di Catania, Via San Nullo 5/i, 95123 Catania, Italy}

\author{M. Frasca}
\affiliation{Dipartimento di Ingegneria Elettrica, Elettronica e dei Sistemi,
Universit\`a degli Studi di Catania, viale A. Doria 6, 95125 Catania, Italy}
\affiliation{Laboratorio sui Sistemi Complessi, Scuola Superiore
di Catania, Via San Nullo 5/i, 95123 Catania, Italy}

\author{J. G\'omez-Garde\~nes}
\affiliation{Departamento de Matem\'atica Aplicada, Universidad Rey Juan Carlos, M\'ostoles 28933, Madrid, Spain}
\affiliation{Institute for Biocomputation and Physics of Complex Systems, University of Zaragoza, 50009 Zaragoza, Spain}
\affiliation{Laboratorio sui Sistemi Complessi, Scuola Superiore
di Catania, Via San Nullo 5/i, 95123 Catania, Italy}

\author{V. Latora}
\affiliation{Dipartimento di Fisica e
    Astronomia, Universit\`a di Catania, and INFN, Via S. Sofia 64, 95123
    Catania, Italy}
\affiliation{Laboratorio sui Sistemi Complessi, Scuola Superiore
di Catania, Via San Nullo 5/i, 95123 Catania, Italy}

\author{Y. Moreno}
\affiliation{Institute for Biocomputation and Physics of Complex Systems, University of Zaragoza, 50009 Zaragoza, Spain}
\affiliation{Department of Theoretical Physics, University of Zaragoza, 50009 Zaragoza, Spain}

\date{\today}

\begin{abstract}
  We address the problem of how the survival of cooperation in a social 
  system depends on the motion of the individuals. Specifically, we study
  a model in which Prisoner's Dilemma players are allowed to move in a
  two-dimensional plane. Our results show that cooperation can survive
  in such a system provided that both the temptation to defect and the
  velocity at which agents move are not too high. Moreover, we show
  that when these conditions are fulfilled, the only asymptotic state
  of the system is that in which all players are cooperators. Our
  results might have implications for the design of cooperative
  strategies in motion coordination and other applications including
  wireless networks.
\end{abstract}


\maketitle

An open question in biology and social sciences is to understand how
cooperation emerges in a population of selfish individuals. 
A theoretical framework that has shed some light into
this long-standing problem is evolutionary game theory
\cite{nowak_book,sf07}. Through the development and the study of
different social dilemmas, scientists have been able to elucidate some
of the mechanisms that enable cooperative behavior in populations. In
particular, one of the most studied games is the Prisoner's Dilemma
(PD), a two-players game in which each individual can only adopt one
of the two available strategies: cooperation (C) or defection
(D). While a population of individuals playing a PD game does not
support cooperation if they are well-mixed, the existence of a spatial
structure gives as a result that cooperation survives under certain
conditions as cooperative clusters can emerge in the system
\cite{nowak_book,sf07}.

In the last years, the field has been spurred by new discoveries on the 
actual structure of the systems to which evolutionary models are 
applied. It turns out that in the vast majority of real-world networks of
interactions \cite{blmch06}, the probability that an individual has $k$ contacts follows a power-law distribution $P(k)\sim
k^{-\gamma}$, being $\gamma$ an exponent that usually lies between 2
and 3. Examples of these so-called scale-free (SF) networks can be
found in almost every field of science \cite{blmch06}. An alternative
to a power-law distribution is a network of contacts that approaches
an exponential tail for $k$ larger than the average connectivity in
the population, being the Erd\"{o}s-Renyi (ER) network the benchmark
of this kind of distribution \cite{blmch06}.

Recent works have shown that cooperative behavior is actually enhanced
when individuals play on complex networks, particularly if the network of contacts is scale-free 
\cite{sp05,gcfm07,pgfm07,vss08}. The reason is that
cooperators are fixed in the highly connected nodes, turning also into
cooperators their neighborhood and guaranteeing in this way their
long-time success. Additionally, several works have explored different
rewiring mechanisms that allow an improvement in the average level of
cooperation in the system \cite{ezcs05,jlcs08,spl06,sp00}. In contrast, cooperation can also be promoted without invoking different rewiring rules \cite{ptn06,vslp09}. Interestingly, social dilemmas can also be used
to generate highly cooperative networks by implementing a growth
mechanism in which the newcomers are attracted to already existing
nodes with a probability that depends on the nodes' benefits
\cite{pgfsm08}.

In spite of the relative large body of work that has been accumulated
in the last few years, there are situations of practical relevance
that remain less explored. This is the case of models where
individuals can move and change their neighborhood continuously by
encountering different game's partners as time goes on. Highly
changing environments can be found in a number of social situations
and the study of how cooperative levels are affected by the inherent
mobility of the system's constituents can shed light on the general
question of how cooperation emerges. Furthermore, the insight gained
can be used to design cooperation-based protocols for communication
between wireless devices such as robots \cite{snm05}. Recently, a few
works have dealt with this kind of situation
\cite{a04,vsa07,hy08,hy09}. However, the models were limited to the case in
which individuals are allowed to move on the sites of a 2D regular
lattice.  In this paper, we consider the less-constrained case in which a set of Prisoner's Dilemma players
unconditionally move on a two dimensional plane. We explore under
which conditions cooperation is sustained. In particular, we inspect
the robustness of the average level of cooperation in the population
under variation of the game parameters and of the mobility rules. Our
results show that cooperation is actually promoted provided that
players do not move too fast and that cooperation is not too
expensive. Additionally, at variance with other cases, the dynamics of
the system exhibits only two stable attractors -those in which the
whole population plays with one of the two possible strategies.


In our model, we consider $N$ agents (individuals) moving in a square
plane of size $L$ with periodic boundary conditions, and playing 
a game on the instantaneous network of contacts. 
The three main ingredients of the model are: the rules of the motion, 
the definition of the graph of interactions, and the rules of the 
evolutionary game.

{\em Motion.} ~~
Each agent moves 
at time $t$ with a velocity ${\bf v}_i(t)$ $(i=1,2,\ldots,N)$.  We assume that individuals can only change their direction of motion, $\theta_i(t)$, but not their speed which  is constant in time, and equal for all the agents.  Hence
we can write the velocities as: ${\bf v}_i(t) = (v \cos \theta_i(t), v
\sin \theta_i(t))$.  The individuals are initially assigned a random
position in the square and a random direction of motion. At each time
step they update their positions and velocity according to the
following dynamical rules:
\begin{equation}
\label{eq:y}
{\bf x}_i(t + 1) = {\bf x_i}(t) + {\bf v_i}(t) 
\end{equation}
\begin{equation}
\label{eq:theta}
\theta_i(t + 1) = \eta_i  
\end{equation}
where ${\bf x}_i(t)$ is the position of the $i$-th agent in the plane
at time $t$ and $\eta_i$ are $N$ independent random variables
chosen at each time with uniform probability in the interval $[-\pi;
\pi]$. 
%
%

{\em Network of interactions.}~~ At each time step we consider that
the neighborhood of a given agent $i$ is made up by all the
individuals $j$ which are within an Euclidean distance $d_{ij}$ less
than some threshold $r$. In what follows, without loss of generality,
we set $r=1$. Therefore, the instant network of contacts is defined as
the graph formed by nodes centered at all the $N$ circles of radius
$1$ together with the links between those agents in the neighborhood
of each individual. Note that as agents move every time step, the
network of contacts, and hence the adjacency matrix of the graph is
continuously changing, not only because the number of contacts an
individual has may change, but also due to the fact that the neighbors
are not always the same. The topological features of the graph defined
above depend on several parameters. For instance, the mean degree of
the graph can be written as $\langle k \rangle = \rho \pi r^2= \rho
\pi $ where $\rho =N/L^2$ is the density of agents. For small values
of $\rho$, the graph is composed by several components and there may
also exist isolated individuals. On the contrary, when $\rho > \rho_c$
a unique giant component appears \cite{dc02} (for our system with
periodic boundary conditions $\rho_c \sim 1.43$).

{\em Evolutionary dynamics}. 
As the rules governing the evolutionary dynamics, we  assume that
individuals interact by playing the Prisoner's Dilemma (PD)
game. Initially, players adopt one of the two available strategies,
namely to cooperate or to defect, with the same probability $1/2$. At
every round of the game all the agents play once with all their
corresponding instant neighbors. The results of a game translate into
the following payoffs: both agents receive $R$ under mutual
cooperation and $P$ under mutual defection, while a cooperator
receives $S$ when confronted to a defector, which in turn receives
$T$. These four payoffs are ordered as $T > R > P \ge S$ in the PD game
so that defection is the best choice, regardless of the opponent
strategy. As usual in recent studies, we choose the PD payoffs as $R =
1$, $ P = S = 0$, and $T = b > 1$.
Once the agents have played with all their neighbors, they accumulate
the payoffs obtained in each game, and depending on their total payoffs
and on the payoffs of the first neighbors, they decide whether or
not to keep playing with the same strategy for the next round
robin. In this process, an agent $i$ picks up at random one of its
neighbors, say $j$, and compare their respective payoffs $P_i$ and
$P_j$. If $P_i > P_j$, nothing happens and $i$ keeps playing with the
same strategy. On the contrary, if $P_j > P_i$ , agent $i$ adopts the
strategy of $j$ with a probability proportional to the payoff
difference:
\begin{equation} 
\Pi_{ij} = \frac{P_j - P_i}{\max\{k_j,k_i\}b}\;, 
\end{equation}
where $k_i$ and $k_j$ are the number of instant neighbors of $i$ and
$j$ respectively ({\em i.e.} the number of agents inside the circles
of radius $r$ centered at $i$ and $j$ respectively). This process of
strategy updating is done synchronously for all the agents of the
system and is a finite population analogue of replicator
dynamics. When finished, the payoffs are reset to zero, so that repeated games are not considered.
 
The movement and game dynamics might in general be correlated, and the
influence of the agents movement on the performance of the PD dynamics
depends on the ratio between their corresponding time scales. Here, we
consider the situation in which both movement and evolutionary
dynamics have the same time scale. Therefore, at each time step, the
following sequence is performed: {\em (i)} the agents perform a new
movement in the two-dimensional space, {\em (ii)} establish the new
network of contacts (determined by the radius $r$ of interaction) and
{\em (iii)} they play a round of the PD game, accumulating the payoffs
and finally updating their corresponding strategies accordingly.
After this latter step, the players move again. The process is
repeated until a stationary state is reached. Here, a stationary state
is one in which no further changes of strategies are possible.


We have performed extensive numerical simulations of the model for
various values of the agent density $\rho$ and velocity $v$, and
different values of the game parameter $b$. Let us first note that for
the limiting case in which $v=0$, the results point out that the
average level of cooperation is different from zero, as one might expect
from the fact that the underlying network of contacts has a 
Poisson degree distribution. 
Indeed, the graph corresponds to a random geometric
graph \cite{dc02}, a network having the same $P(k)$ as an ER random 
graph, but with a higher clustering coefficient. This latter feature
leads to a further increment of the average level of cooperation, as
it has been shown that a network with a high clustering coefficient
promotes cooperation \cite{pwp08,agl08}.

\begin{figure}[t]
\begin{center}
\includegraphics[width=\columnwidth]{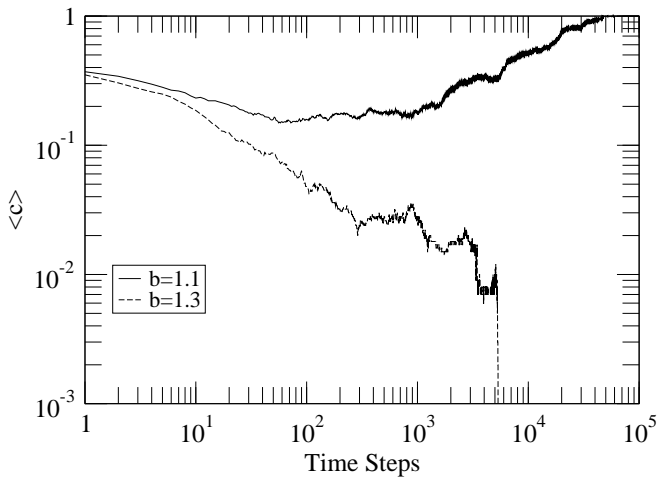}
\caption{Average level of cooperation,  $\langle c \rangle$, as a function of time (Monte Carlo steps) for $v=0.01$ and two different values of $b$, $b=1.1$ $b=1.3$, as indicated. Other model parameters have been fixed to $\rho=1.30$ and $N=10^3$ agents.} \label{fig1}
\end{center}
\end{figure}

Let us now focus on the case $v \neq 0$. The first difference that
arises with respect to the case in which agents do not move is that
the dynamics of the system only have two attractors. Namely, the
asymptotic state (i.e., when the probability that any player changes
its strategy is zero) is either a fully cooperative network (all-C) or
a network in which all the individuals end up playing as defectors
(all-D). This behavior is illustrated in Fig.\ \ref{fig1}, where we
have reported the average level of cooperation $\langle c \rangle$ in
a population of $N=10^3$ individuals as a function of time, for
$v=0.01$ and for two different values of $b$. Starting from a
configuration in which individuals are cooperators or defectors with the same probability, the average level of cooperation slowly evolves to
one of the two asymptotic states: all-C or all-D. It is also worth
stressing that the system reaches those states more slowly than in
static settings (i.e., when $v=0$). Specifically, it appears that the
system spends a considerable time in metastable states (flat regions
in the figure) that are followed by a sudden decrease (or increase) of
the average level of cooperation.

\begin{figure}[t]
\begin{center}
\includegraphics[width=\columnwidth]{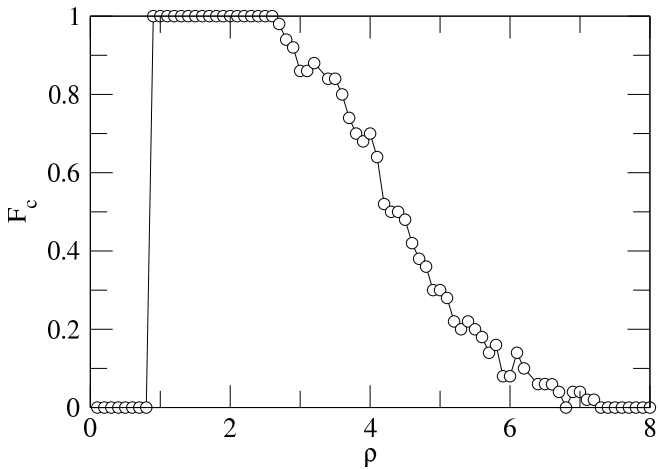}
\caption{Fraction of realizations in which the system ends up in an all-C configuration, $F_c$, as a function of the density of players $\rho$ for a fixed value of $b=1.1$ and $v=0.01$. The system is made up of $N=10^3$ agents. The results are averages taken over 100 different realizations.}
\label{fig2}
\end{center}
\end{figure}

The evolution of the system depends on the density of players. In Fig.\ \ref{fig2}, we have represented the dependence of the fraction of realizations, $F_c$ in which the population ends up in an all-C configuration as a function of the density $\rho$ for $b=1.1$ and $v=0.01$. There are two limits for which $F_c=0$. At low values of the density, the agents are too spread in the 2D plane. As a result, cooperators unsuccessfully strive to survive and get extinguished given the low chance they have to form clusters -the only mechanism that can enforce their success. On the contrary, for large values of $\rho$ the population is quite dense and, locally, the agents' neighborhoods resemble a well-mixed population in which more or less everybody interacts with everybody and therefore defection is the only possible asymptotic state. Values of $\rho$ between these two limiting cases confer to cooperators a chance to survive. Interestingly, there is a region of the density of players, $0.9 \lesssim \rho \lesssim 3$ which is optimal for cooperative behavior. Beyond this region $F_c$ decays exponentially with $\rho$ reaching zero at $\rho\approx 7$.

\begin{figure}[t]
\begin{center}
\includegraphics[width=2.0in,angle=-90]{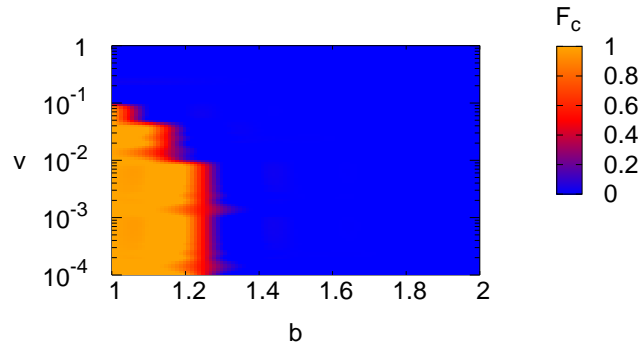}
\caption{(Color online) The color code shows the fraction of realizations in which the whole system is made up of cooperators, $F_c$, as a function of the velocity at which the agents move ($v$) and the temptation to defect ($b$). The Y-axis is in log scale for clarity. The rest of parameters are $N=10^3$ agents and $\rho=1.30$. Each point is an average over 100 different realizations} \label{fig3}
\end{center}
\end{figure}

Up to now, we have analyzed the behavior of the system for small values of the velocity of the agents and of the temptation to defect. Figure\ \ref{fig3} summarizes the results obtained for a wider range of model parameters ($v$ and $b$) in a population of $N=10^3$ agents and $\rho=1.3$. The results are averages taken over 100 realizations of the model. The phase diagram shows a relative wide region of the model parameters in which cooperative behavior survives. For a fixed value of $v$, this region is bounded by a maximum value of the temptation to defect close to $b=1.3$, which decreases as the velocity at which players move increases. Furthermore, when $b$ is kept fixed, increasing the value of $v$ is not always beneficial for the survival of cooperation. In fact, when the individuals move too fast, they change their environment quite often and quickly, then increasing the likelihood to meet each time step a completely different set of players. In other words, when the velocity is increased beyond a certain value, the well-mixed hypothesis applies to the whole population of players, thus leading to the extinction of cooperation in the long time limit.

\begin{figure}[!t]
\begin{center}
\includegraphics[width=0.90\columnwidth]{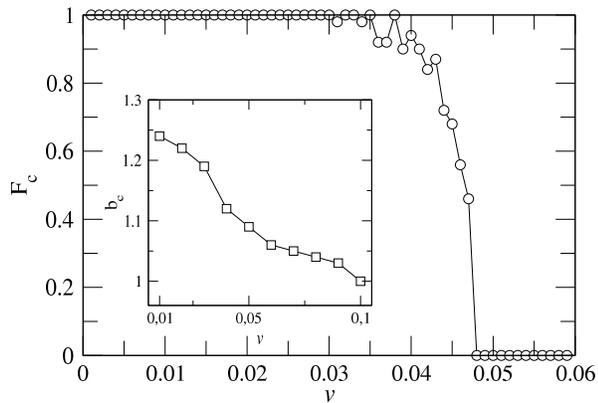}
\caption{Fraction of realizations ending up in an all-C configuration as a function of the velocity $v$ of the agents for $b=1.1$. The inset shows the smallest value of the temptation to defect, $b_c$, for which the probability of achieving a fully cooperator asymptotic state is zero, as a function of $v$. In both cases, $N=10^3$ agents, $\rho=1.30$, and results correspond to averages over 100 realizations.} \label{fig4}
\end{center}
\end{figure}

Figure\ \ref{fig4} sheds more light on the dependence of the fraction of cooperators with respect to the velocity of the agents. There we have represented the layer corresponding to $b=1.1$ in Fig.\ \ref{fig3}. As can be seen from the figure, for low values of $v$ all the realizations lead the system to a configuration in which all strategists are cooperators. As the PD players move faster, the probability of achieving such a configuration decreases and gets zero for values of $v$ close to $0.05$. From that point on, the all-C asymptotic state is never realized. This latter point also depends on the specific value of $b$. The inset of Fig.\ \ref{fig4}, represents the smallest values of the temptation to defect, $b_c$, for which in all the realizations performed the system ended up in the all defectors state as a function of $v$. The results show that beyond $v\approx 0.1$, cooperation never survives in a population of moving agents irrespective of $b$.


In short, we have studied the effects of mobility on a population of Prisoner's Dilemma players that are able to move in a two-dimensional plane. Numerical simulations of the model show that a fully cooperative system is sustained when both the temptation to defect and the velocity of the agents are not too high. Although cooperation is extinguished for a wide region of the parameter space, our results show that mobility have a positive effect on the emergence of cooperation. As a matter of fact, as soon as $v \neq 0$, the mobility of the agents provokes the spread of the winning strategy to the whole population, leading the system to a global attractor in which all players share the surviving strategy. In other words, the movement of individuals prevents the coexistence of different strategies in the long time limit. Namely, for small (and fixed) values of $b$ cooperation prevails at low velocities, while defection succeeds for larger $v$. Our results are relevant for the design of new cooperation-based protocols aimed at motion coordination among wireless devices and for other communication processes based on game theoretical models \cite{snm05}.

\begin{acknowledgments}
 Y. M. is supported by MCINN through the Ram\'{o}n y Cajal
  Program. This work has been partially supported by
  the Spanish DGICYT Projects FIS2006-12781-C02-01, and FIS2008-01240,
  and a DGA grant to FENOL group.
\end{acknowledgments}


\begin{thebibliography}{99}

\bibitem{nowak_book} M. A. Nowak, {\em Evolutionary Dynamics:
Exploring the Equations of Life}. (Harvard University Press,
Cambridge, Massachusetts, and London, England, 2006).

\bibitem{sf07} G. Szab\'o, and G. F\'ath, \emph{Phys. Rep.} {\bf 446}, 97 (2007).

\bibitem{blmch06} S. Boccaletti, V. Latora, Y. Moreno, M. Chavez, and D.-U.
  Hwang, Phys. Rep. {\bf 424}, 175 (2006).
  
\bibitem{sp05} F.C. Santos and J. M. Pacheco, Phys. Rev. Lett.  {\bf 95}, 098104 (2005).

\bibitem{gcfm07} J. G\'{o}mez-Garde\~{n}es, M. Campillo, L. M. Flor\'{\i}a and Y. Moreno,  Phys. Rev. Lett {\bf 98}, 108103
(2007).

\bibitem{pgfm07} J. Poncela, J. G\'omez-Garde\~nes, L. M. Flor\'{\i}a, and Y. Moreno, New J. Phys. {\bf 9}, 184 (2007).

\bibitem{vss08} J. Vukov, G. Szab\'o, and A. Szolnoki, Phys. Rev. E {\bf 77}, 026109 (2008).

\bibitem{ezcs05} V. M. Egu\'{\i}luz, M. G. Zimmermann, C. J. Cela-Conde, and M. San  Miguel, Am. J. Soc. \textbf{110}, 977 (2005).
  
\bibitem{jlcs08} R. Jim\'enez, H. Lugo, J. A. Cuesta, and A. S\'anchez, J. Theor. Biol. {\bf 250}, 475 (2008). 

\bibitem{spl06} F. C. Santos, J. M. Pacheco, T. Lenaerts,  PLoS Comput. Biol. {\bf 2}(10), e140 (2006).

\bibitem{sp00} B. Skyrms and R. Pemantle, Proc. Natl. Acad. Sci. USA {\bf 97}, 9340 (2000).

\bibitem{ptn06} J. M. Pacheco, A. Traulsen, and M. A. Nowak, Phys. Rev. Lett. {\bf 97} 258103 (2006).

\bibitem{vslp09} S. Van Segbroeck, F. C. Santos, T. Lenaerts, and J. M. Pacheco, Phys. Rev. Lett. {\bf 102}, 058105 (2009).

\bibitem{pgfsm08} J. Poncela, J. G\'omez-Garde\~nes, L. M. Flor\'{\i}a, A. S\'anchez and Y. Moreno, PLoS ONE {\bf 3}, e2449 (2008).

\bibitem{snm05} V. Srivastava, J. Neel, A. B. MacKenzie, \emph{et al.}, IEEE Communications {\bf 7}, 46 (2005).

\bibitem{a04} C. A. Aktipis, J. Theor. Biol. {\bf 231}, 249 (2004). 

\bibitem{vsa07} M. H. Vainstein, A. T. C. Silva, and J. J. Arenzon, J. Theor. Biol.  {\bf 244}, 722 (2007).

\bibitem{hy08} D. Helbing and W. Yu, Adv. Comp. Syst. {\bf 11}, 641 (2008).

\bibitem{hy09} D. Helbing and W. Yu, Proc. Nat. Acad. Sci. USA {\bf 106}, 3680 (2009).

\bibitem{dc02} J. Dall, and M. Christensen, Phys. Rev. E {\bf 66}, 016121 (2002).

\bibitem{pwp08} A. Pusch, S. Weber, and M. Porto, Phys. Rev. E {\bf 77}, 036120 (2008).

\bibitem{agl08} S. Assenza, J. G\'omez-Garde\~nes, and V. Latora, Phys. Rev. E {\bf 78}, 017101 (2008).

\end{thebibliography}
\end{document}